\documentclass[aps,prl,twocolumn,superscriptaddress]{revtex4-2}

\usepackage{graphicx}% Include figure files
\usepackage{dcolumn}% Align table columns on decimal point
\usepackage{bm}% bold math
\usepackage{hyperref}% add hypertext capabilities
\usepackage[mathlines]{lineno}% Enable numbering of text and display math
\usepackage{braket}
\usepackage{amsmath}
\usepackage{amssymb}
\begin{document}

\title{Geometric representation of higher-order optical modes}

\author{Claire Cisowski}
%\email[]{Your e-mail address}
\affiliation{School of Physics and Astronomy, University of Glasgow, G12 8QQ, Glasgow, UK.}

\date{\today}

\begin{abstract}
An octant representation of higher-order optical modes that includes Laguerre-Gaussian and Hermite-Gaussian modes is presented. The octant picture captures the high-dimensional nature of three-state optical systems and beyond, with standard Poincar\'e spheres for orbital angular momentum forming subspaces of the entire state space. 
This representation enables intuitive manipulation of both classical modes and optical qudits and provides a framework for extending Berry phases and topological invariants to high dimensions.
\end{abstract}

% insert suggested keywords - APS authors don't need to do this
%\keywords{}

\maketitle

The ability to form mental pictures of physical systems has shaped the history of physics for centuries, with two prominent examples being the Bloch sphere representation of qubits in quantum mechanics~\cite{Bloch1946} and the Poincar\'e sphere representation of fully polarized light in optics~\cite{poincare1892}. Recently, the exploration of structured light modes, including Hermite-Gaussian modes $HG_{nm}$ modes of indices $n$ and $m$ and Laguerre-Gaussian modes $LG_p^\ell$ of radial index $p$ and azimuthal index $\ell$, has motivated the development of a geometric toolkit for spatial light. 
Poincar\'e sphere analogues have been introduced to describe  
first order modes~\cite{Padgett:99}, with the mode order being defined as $N=n+m=2p+|\ell|$, and to represent vector vortex modes in~\cite{Milione2011}. 
By accessing large Hilbert spaces, 
higher order spatial modes ($N\geq 1$) have emerged as a promising resource for high-dimensional quantum communication~\cite{Mirhosseini_2015}. However, visualizing their state spaces is challenging; spheres are no longer sufficient to describe the entire state space, and hyperspheres must be used instead~\cite{Bengtsson_Zyczkowski_2006}.  
It is possible to explore subspaces using Poincar\'e sphere analogues and $SU(2)$ Lie algebra~\cite{Calvo:05}, but capturing the topological properties of the entire state space is valuable for studying geometric phases of light in high dimensions~\cite {ciso2022}, holonomic computational gates and minimal length paths for quantum communication applications~\cite{doi:10.1126/science.1121541,Brody:1999cw}.

In this Letter, it is shown that the state space of higher-order optical modes can be conveniently visualized as the positive octant of a sphere spanned by a torus. The case of second-order modes is taken as a representative example. In contrast to Majorana's stellar representation~\cite{Majorana1932}, which strength lies in the identification of symmetries in high-dimensional systems, the octant picture directly encodes relative amplitudes and phase between the three base states within its coordinates~\cite{Bengtsson_Zyczkowski_2006}. Octant coordinates are well adapted to the Fubini-Study metric~\cite{Bengtsson_Zyczkowski_2006}, providing a natural way to measure transition probabilities, distinguishability, and quantum information~\cite{FACCHI20104801}.   

Second-order optical modes form a three-state system, written in the formalism of quantum mechanics as $\ket{\psi}=(\psi_1,\psi_2,\psi_3)$, with $\psi_i$ denoting complex mode amplitudes. Normalization $|\psi_1^2|+|\psi_2|^2+|\psi_3|^2=1$ reduces the number of parameters needed to describe the system from six real numbers to five, defining an hypersphere $S^5$. The space of pure states emerges by fixing an overall phase under the equivalence relation $\ket{\Psi}=\lambda\ket{\Psi}$ where $\lambda$ is any non-zero complex number. It corresponds to the complex projective space $\mathbb{C}\text{P}^2=S^5/U(1)$. Four independent parameters are therefore needed to parametrize the state space of second-order optical modes. The octant representation introduces four local angle type variables, $(\theta,\phi,\chi_1,\chi_2)$, to cover most of $\mathbb{C}\text{P}^2$~\cite{Bengtsson:2001yd}. Indeed, just as two coordinate charts are needed to cover a sphere, three charts are required to cover the entire space~\cite{Nakahara:2003}.  
Omitting $\psi_3=0$:
\begin{multline}\label{eq.octant}
\ket{\psi} = e^{i\delta} \Big(
\sin\theta \cos\phi \, e^{i\chi_1} \ket{\psi_1} \\
+ \sin\theta \sin\phi \, e^{i\chi_2} \ket{\psi_2} 
+ \cos\theta \, \ket{\psi_{3}}
\Big),
\end{multline}
where $e^{i\delta}$ is an overall phase factor, $0\leq\!\theta\!<\!\pi/2$ and $0\!\leq\!\phi\,\leq\!\pi/2$ set the magnitudes of the base states and $0\,\leq\,\chi_{1,2}\,<\,2\pi$ determine their relative phases with $\chi_{1}=\rm{arg}(\psi_1)-\rm{arg}(\psi_3)$ and $\chi_2=\rm{arg}(\psi_2)-\rm{arg}(\psi_3)$. 
The locus $\psi_3=0$ is a coordinate singularity and forms a sphere, generalizing point-like singularities in two-state systems~\cite{Nakahara:2003}. 
% at  \begin{equation}\label{eq.param}
%(\psi_1,\psi_2,\psi_3) =e^{i\delta} (n_0 e^{i\chi_1},n_1 %e^{i\chi_2},n_2)   
%\end{equation}
%where $n_0,n_1,n_2\geq 0$ and $n_0^2+n_1^2+n_2^2=1$ (notmalized state vectors) we can use $n_i$ as orthographc coordinates or we can set $y_i=n_i^2$ and then we have $y_0+y_1+y_2=1$ 
Eq.~\ref{eq.octant} holds for any three orthonormal basis $\{ \ket{\psi_1},\ket{\psi_2},\ket{\psi_3}\}$. 
In each such basis, points on the octant surface can be written as a superposition of all three base states, the edges of the octant correspond to two-state superpositions, and its corners identify base states. Allowing $\chi_1$ and $\chi_2$ to vary from $0$ to $2\pi$ generates a torus above each point of the octant. This torus collapses to a circle on the edges and to a point at the corners where $\chi_1$ and/or $\chi_2$ become undefined. This is readily illustrated with standard second-order structured light modes. 

\begin{figure}[h]
\includegraphics[width=\columnwidth]{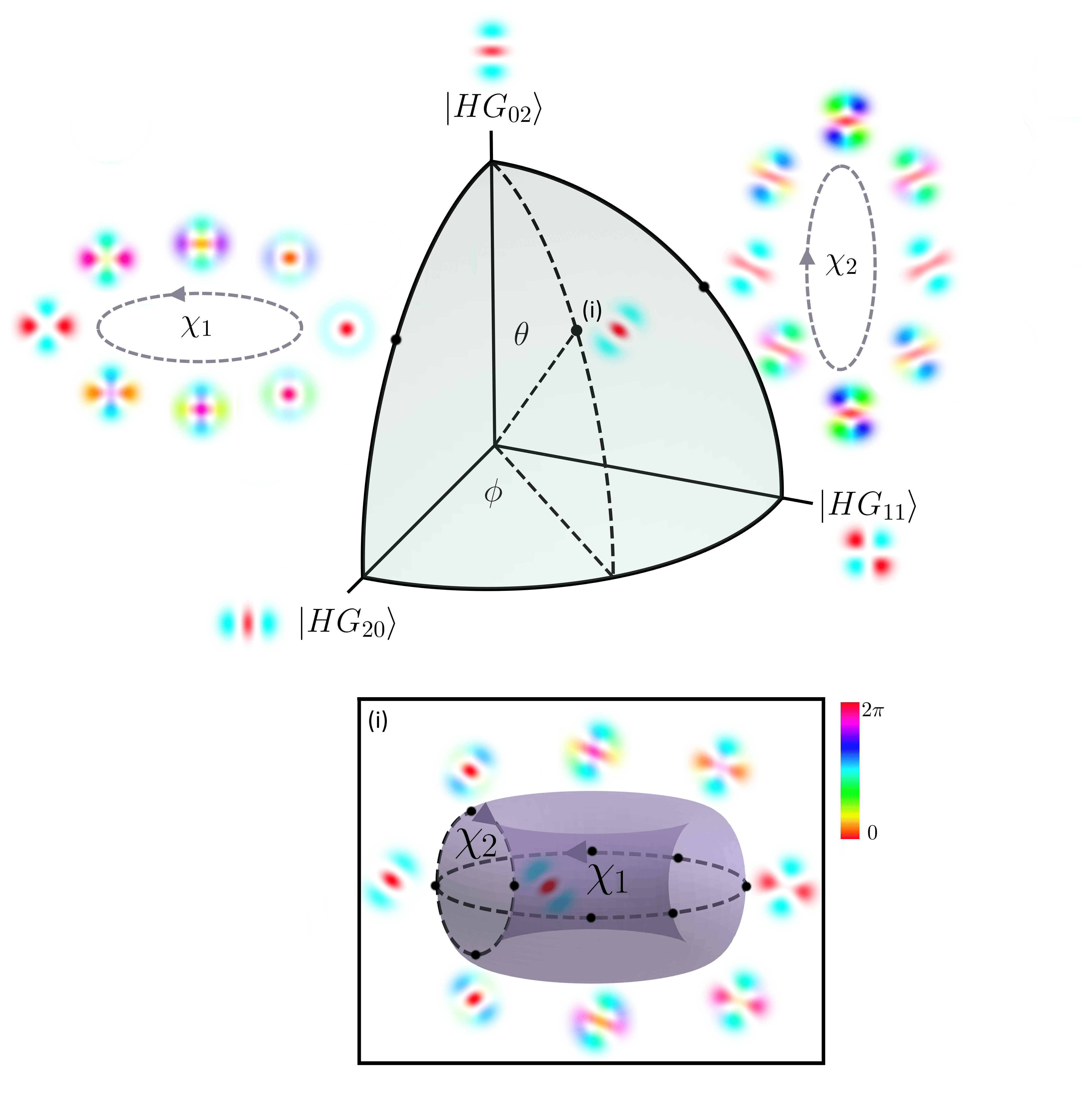}
\caption{\label{fig:HGbasis} Octant representation of second order HG modes. Intensity modulated phase distributions are provided for a set of states. The inset shows the torus structure above (i).}
\end{figure}

In the basis of Hermite-Gaussian (HG) modes, Eq.~\ref{eq.octant} becomes:
\begin{multline}\label{eq.HGbasis}
\ket{\psi} = e^{i\delta} \Big(
\sin\theta \cos\phi \, e^{i\chi_1} \ket{\text{HG}_{20}} \\
+ \sin\theta \sin\phi \, e^{i\chi_2} \ket{\text{HG}_{11}} 
+ \cos\theta \, \ket{\text{HG}_{02}}
\Big)
\end{multline}
Fig.~\ref{fig:HGbasis} shows the octant representation of HG modes corresponding to Eq.~\ref{eq.HGbasis}. The corners identify the base modes  $\text{HG}_{20}$, $\text{HG}_{11}$ and $\text{HG}_{02}$. Along the left edge, states are superpositions of $\text{HG}_{02}$ and $\text{HG}_{20}$ with $\chi_2$ undefined. One such state, of coordinates (${\theta=\pi/4,\phi=0,\chi_1=0,\chi_2=0}$), is the Laguerre-Gaussian (LG) mode $\text{LG}_1^0$: 
\begin{equation}
\ket{\text{LG}_1^{0}} =\frac{1}{\sqrt{2}}\ket{\text{HG}_{20}}+\frac{1}{\sqrt{2}}\ket{\text{HG}_{02}}. 
\end{equation}
The right edge of the octant contains superpositions of $\text{HG}_{02}$ and $\text{HG}_{11}$ modes with $\chi_1$ undefined.
Each edge forms a subspace of $\mathbb{C}\text{P}^2$ and can be mapped onto a modal Poincar\'e sphere by taking the states at the two corners as the poles and identifying the mid-point state and its phase variations as points spanning the equator. It is worth noting that these subspaces stand out due to a choice of basis and, as such, carry no special significance over modal spheres available in alternative basis from a topological point of view.

The octant representation clearly illustrates the action of general mode-preserving converters in the HG basis. Building on the formalism of O'Neil and Courtial~\cite{ONEIL200035}, the $SU(3)$ transformation matrix of a $N=2$ mode converter, also known as $\varphi$-converter, can be expressed as:
\begin{equation}\label{modeconverter}
C_{\text{HG}}(\varphi)=e^{i2\varphi}
\begin{pmatrix}
e^{-i2\varphi} & 0 & 0 \\
0 & e^{-i\varphi} & 0 \\
0 & 0 & 1 
\end{pmatrix},   
\end{equation}
in the HG basis.
Where the overall phase factor $e^{-i2\varphi}$ can be discarded. The relative mode amplitudes are not affected by $C_{\text{HG}}(\varphi)$; the $(\theta,\phi)$ coordinates are unchanged, the state remains at its position on the surface of the octant. Its position on the torus, however, varies. This is made explicit by identifying $\chi_1=-2\varphi$ and $\chi_2=-\varphi$.     
Experimentally, $\varphi$-converter are typically realized using a pair of cylindrical lenses~\cite{BEIJERSBERGEN1993123}, with the $\varphi=\pi$ and the $\varphi=\pi/2$ configurations being the most commonly employed. The $\pi$-converter transforms a mode with orbital angular momentum $-\ell$ to one with $-\ell$ and corresponds to an image reflection. 
In the octant picture, the $\pi$-converter takes a state to the opposite side of the circle formed by $\chi_2$ as $\chi_1=-2\pi$ and $\chi_2=-\pi$. This can be seen in Fig.~\ref{fig:HGbasis} where the mirror image of the state ($\theta=\pi/4,\phi=\pi/2,\chi_1=undefined,\chi_2=0$), lies at the opposite side of the $\chi_2$ circle. The $\pi/2$-converter can transform a mode with no orbital angular momentum into one possessing it~\cite{BEIJERSBERGEN1993123}. In the octant representation, the trajectory induced on the torus is given by $\chi_1=-\pi$, $\chi_2=-\pi/2$. Consequently, in the HG basis, $\text{HG}_{20}$ modes rotated at $\pm 45^{\circ}$ with respect to the $x$-axis, $\text{LG}_0^2$ and $\text{LG}_0^{-2}$ modes lie on the same torus above the point $\theta\approx 0.333\pi, \phi\approx0.304\pi$.    

%\begin{equation}
%\ket{LG_0^2} =\frac{1}{2}\ket{HG_{20}}+\frac{i}{\sqrt{2}}\ket{HG_{11}}-\frac{1}{2}\ket{HG_{02}} 
%\end{equation}
%\begin{equation}
%\ket{LG_0^{-2}} =\frac{1}{2}\ket{HG_{20}}-\frac{i}{\sqrt{2}}\ket{HG_{11}}-\frac{1}{2}\ket{HG_{02}} 
%\end{equation}
%\begin{equation}
%\ket{HG_{20}(\Delta=45)} =\frac{1}{2}\ket{HG_{20}}-%\frac{1}{\sqrt{2}}\ket{HG_{11}}-\frac{1}{2}\ket{HG_{02}} 
%\end{equation}

\begin{figure}[h]
\includegraphics[width=\columnwidth]{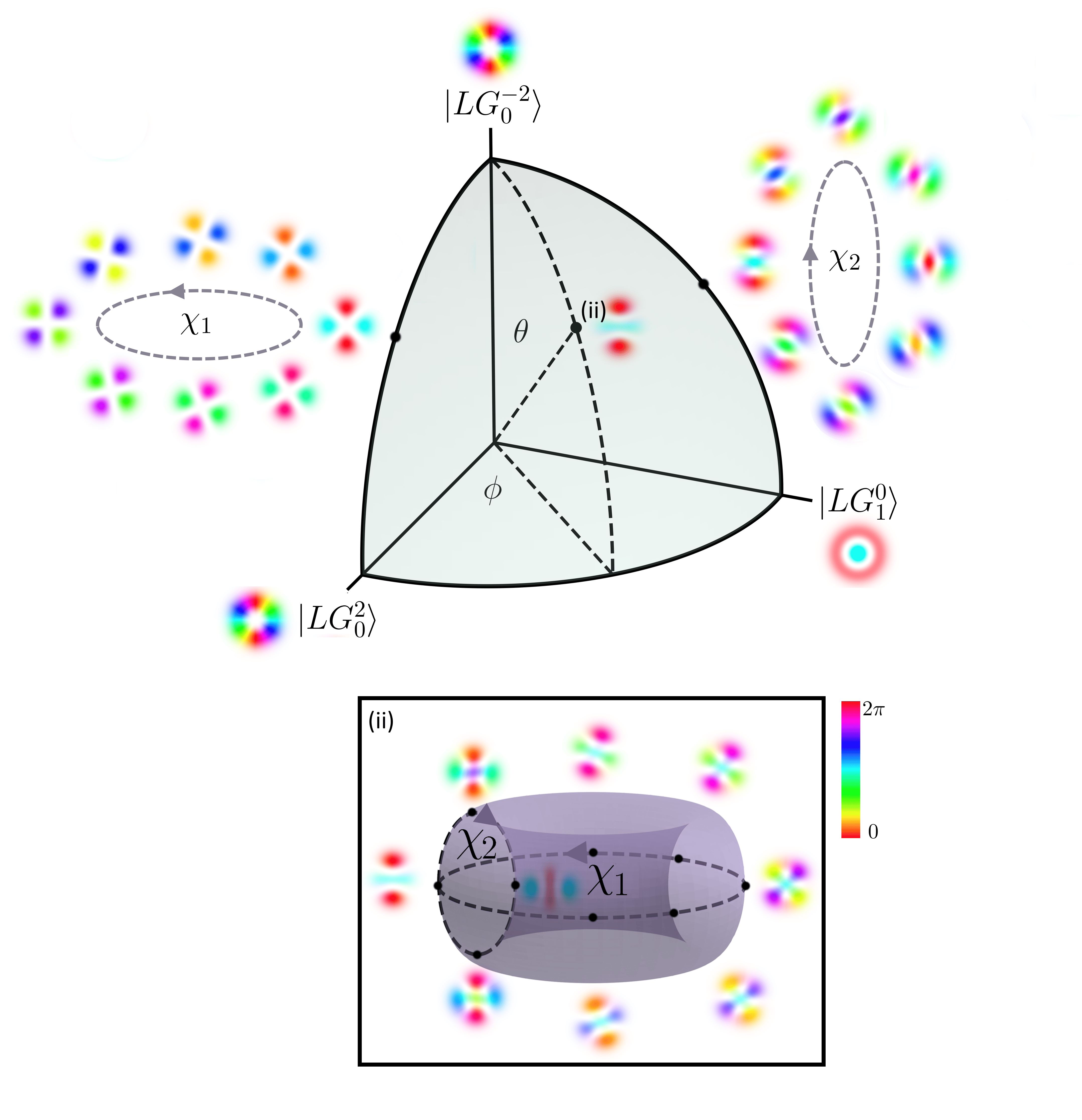}
\caption{\label{fig:LGbasis} Octant representation of second order LG modes. Intensity modulated phase distributions are provided for a set of states. The inset shows the torus structure above (ii).
}
\end{figure}

In addition to mode converters, image rotators constitute another important class of transformations for HG and LG modes. In the HG basis, a rotation will mix the relative mode amplitudes~\cite{ONEIL200035}, forming intricate paths on the octant involving a change in all parameters.   
The LG basis provides a more natural setting for describing the action of image rotators. In the LG basis, Eq.~\ref{eq.octant} becomes: 
\begin{multline}\label{eq.LGbasis}
\ket{\psi} = e^{i\delta} \Big(
\sin\theta \cos\phi \, e^{i\chi_1} \ket{{\text{LG}_{0}^2}} \\
+ \sin\theta \sin\phi \, e^{i\chi_2} \ket{\text{LG}_{1}^0} 
+ \cos\theta \, \ket{\text{LG}_{0}^{-2}}
\Big).
\end{multline}
Fig.~\ref{fig:LGbasis} shows the octant representation of second-order LG modes.  
The edges can be mapped onto Poincar\'e spheres of orbital angular momentum. The right edge defines superpositions of $\text{LG}_1^0$ and $\text{LG}_0^{-2}$ modes, whereas the left edge corresponds to superpositions of $\text{LG}_0^2$ and $\text{LG}_0^{-2}$ modes, such as the one presented in~\cite{Habraken2010}.
It is important to note that, unlike the sphere of first-order modes, the states at the equator of the sphere obtained by taking $\text{LG}_0^{\pm 2}$ as the poles cannot simply be obtained through the action of a cylindrical lens mode converter.

\begin{figure}[h]
\includegraphics[width=\columnwidth]{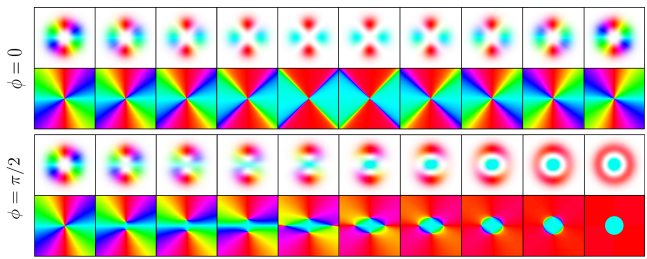}
\caption{\label{fig:leftedge} Phase profile of second-order LG modes, with and without intensity modulation, for increasing $\theta$ and fixed $\phi$ (in all cases $\chi_1=\chi_2=0$).}
\end{figure}

The octant representation brings insights into optical vortex dynamics by mapping smooth transitions between different weighted superpositions of orbital angular momentum eigenmodes. Fig.~\ref{fig:leftedge} shows the evolution of the phase profile on the octant edges. Handedness reversal occurs along the left edge ($\phi=0$), mediated by a state carrying an average of zero orbital angular momentum. The right edge  ($\phi=\pi/2$) illustrates the transition from a beam carrying an orbital angular momentum $\ell=+2$ to one with $\ell=0$, evidencing the splitting of the vortex into two phase singularities of topological charge $\ell=+1$, which are subsequently annihilated by two vortices of opposite charge migrating from spatial infinity. Two mechanisms are therefore highlighted, complementing prior studies~\cite{Molina-Terriza_2012}.    

In the LG basis, the $SU(3)$ mode rotator matrix is~\cite{ONEIL200035}:
\begin{equation}\label{moderotator}
R_{LG}(\Delta)=e^{-i2\Delta} 
\begin{pmatrix}
e^{i4\Delta} & 0 & 0 \\
0 & e^{i2\Delta} & 0 \\
0 & 0 & 1
\end{pmatrix},   
\end{equation}
where $\Delta$ is the rotation angle. Eq.\ref{moderotator} confines the state to the torus, without changing its angular position on the octant surface, with $\chi_1=4\Delta$ and $\chi_2=2\Delta$. A $45^{\circ}$ rotation ($\Delta=\pi/4$) will therefore impose a coordinate change of $\chi_1=\pi$ and $\chi_2=\pi/2$, which can clearly be seen in Fig.~\ref{fig:LGbasis} where the point ($\theta=\pi/4,\phi=0,\chi_1=0,\chi_2=undefined$) is taken to the opposite side of the $\chi_1$ circle and for the point ($\theta=\pi/4,\phi=\pi/2,\chi_1=undefined,\chi_2=0$ which travels by a quarter of circle. $\varphi$-mode converters on the other hand, produce intricate paths in the LG basis, mixing the mode amplitudes and varying relative phases altogether~\cite{ONEIL200035}.

The octant representation provides a clear picture of the entire state space of second-order optical modes, embedding Poincar\'e-like modal spheres within a broader framework and revealing the geometric action of mode converters and mode rotators. It also provides a foundation for exploring vortex dynamics in high-order systems.
Its generalization to higher dimensions is straightforward. For a mode of order $N$, the state space $\mathbb{C}\text{P}^{N-1}$ can be represented by positive hyperoctant of an $N\!-\!1$-sphere spanned by an $N\!-\!1$-torus ~\cite{Bengtsson_Zyczkowski_2006}. 
The states are parametrized both by $N$ phase angles $0\leq \chi_i <2\pi$ describing the torus and by local angles $0<\vartheta_i<\pi/2$ describing the geometry of the hyperoctant. 
Rearranging terms for clarity, Eq.\ref{eq.octant} becomes:   
\begin{multline}\label{eq.octantgen}
\ket{\psi} = e^{i\delta} \Big(
n_0 \ket{\psi_1} + n_1 e^{i\chi_1} \ket{\psi_2} 
+...+ n_{N}e^{i\chi_N}\ket{\psi_{N+1}}
\Big),
\end{multline}
where
\begin{equation}\label{eq.octantpara}
\begin{aligned}
n_0 &= \cos\vartheta_1 \sin\theta_2 \sin\vartheta_3 \cdots \sin\vartheta_n , \\
n_1 &= \sin\vartheta_1 \sin\vartheta_2 \sin\vartheta_3 \cdots \sin\vartheta_n , \\
n_2 &= \cos\vartheta_2 \sin\vartheta_3 \cdots \sin\vartheta_n , \\
&\;\;\vdots  \\
n_n &= \cos\vartheta_n .
\end{aligned}
\end{equation}
Information of relative mode amplitudes and phases are therefore elegantly encoded into two complementary geometric structures, the torus and the hyperoctant.  

The octant representation provides a new tool for studying quantum information theory, phase holonomies and topological invariants in high-dimensional structured light ~\cite{Habraken2010,Galvez2005,VANENK199359}. While this letter focuses on scalar modes, a vectorial extension of the octant representation is only a step away, with the potential to highlight new families of vector vortex modes and introduce additional geometric features arising from the nature of the composite system itself.
The strength of geometric models is their broad applicability. As a universal description of (pure) high-dimensional state systems, the octant representation can facilitate interdisciplinary knowledge transfer through a shared geometric language. Arguably, one of the most recent example of such exchange mediated by a geometric model is that of optical skyrmions, which emerged as geometric counterparts of magnetic skyrmions~\cite{cisorat,10.1098/rspa.2024.0109}.  
Exploring $\mathbb{C}\text{P}^N$ through simple optical transformations could even bring insights into gravitational models~\cite{Gibbons1978}.

\begin{acknowledgments}
The author is grateful to Prof. Brendan Owens for insightful discussions on projective spaces. Financial support from an Early Career Leverhulme fellowship ECF-2023-099 is also acknowledged, in addition to a Lord Kelvin Adam Smith Fellowship  LKAS-20318 and an academic returner support fund from the University of Glasgow.    
\end{acknowledgments}

% Create the reference section using BibTeX:
\bibliography{apssamp}

\end{document}